\documentclass[11pt,article]{aastex}
\usepackage{emulateapj5}
\usepackage{graphicx}
\usepackage{grffile}
\usepackage{natbib}

\shorttitle{The Contribution of Supernovae to Reionization}
\shortauthors{J.~L. Johnson and S. Khochfar}

\begin{document}

\title{The Contribution of Supernovae to Cosmic Reionization}

\author{Jarrett L. Johnson\altaffilmark{1,2} and Sadegh Khochfar\altaffilmark{1}}

\altaffiltext{1}{Max-Planck-Institut f{\"u}r extraterrestrische Physik, 
Giessenbachstra\ss{}e, 85748 Garching, Germany \\
Theoretical Modeling of Cosmic Structures Group; jjohnson@mpe.mpg.de}
\altaffiltext{2}{Los Alamos National Laboratory, 
Los Alamos, NM  87545, USA  \\
Nuclear and Particle Physics, Astrophysics and Cosmology Group (T-2)}


\begin{abstract}
While stars are widely discussed as the source of the high energy photons which reionized the universe, an additional source of ionizing photons that must also contribute to reionization in this 
scenario is the supernovae (SNe) which mark the end of the life of massive stars.  Here we estimate the relative contributions of SNe and stars 
to reionization.  While the rate at which ionizing photons are produced in SN shocks is well below that at which they are produced by stars, the harder spectra of radiation
emitted from SNe leads to an enhanced escape fraction of SN generated photons relative to that of stellar photons.
In particular, along a given line of sight out of a galaxy, we find that for neutral hydrogen column densities $N_{\rm H}$ $\ga$ 10$^{18}$ cm$^{-2}$ the contribution to reionization from 
SNe is greater than that from stars. Drawing on the results of simulations presented in the literature, we find that the overall (line of sight-averaged) SNe shock-generated ionizing 
photon escape fraction is larger than the stellar photon escape fraction by a factor of $\simeq$ 4 to $\simeq$ 7, depending on the metallicity of the stellar population.  Overall, our results 
suggest that the effect of SNe is an enhancement of up to $\sim$ 10 percent in the fraction of hydrogen reionized by stellar sources.  We briefly discuss the 
implications of our results for the population of galaxies responsible for reionization.
\end{abstract}


\section{Introduction}
The reionization of the universe was one of the paramount milestones in the transition from the cosmic Dark Ages, the period before the formation of the first stars, to the present day (see e.g. Barkana \& Loeb 2001; Ciardi \& Ferrara 2005).  As it is the first generations of luminous objects that were the sources of the high energy radiation which reionized the intergalactic medium (IGM), theoretical studies of how reionization took place (e.g. Gnedin \& Ostriker 1997; Miralda-Escud{\' e} et al. 2000; Iliev et al. 2006; McQuinn et al. 2007; Shin et al. 2008) in conjunction with observations of the radiation emitted during the epoch of reionization (e.g. Fan et al. 2006; Robertson et al. 2010; Bowman \& Rogers 2010) stand to teach us a great deal about the nature of the first stars and galaxies (see e.g. Bromm \& Yoshida 2011).

One of the outstanding questions pertaining to reionization is what were the sources of the ionizing radiation which effected it.  It appears unlikely that the progenitors of the galaxies observed to date at $z$ $\ga$ 7 are responsible for reionizing the universe, as the escape fraction of ionizing photons from these galaxies would have to be very large or, otherwise, a low-metallicity stellar population or one with a relatively top-heavy mass function would be required (see e.g. Ouchi et al. 2009; Bunker et al. 2010; Wilkins et al. 2011; but see also Wyithe et al. 2009; Bouwens et al. 2011).  However, there are a number of possible sources which likely contributed to reionization at higher redshifts, where observations of the cosmic microwave background (CMB) suggest a large fraction of the IGM was reionized (Komatsu et al. 2009).  
Among these are stellar sources, including possibly massive Population (Pop) III stars (e.g. Ciardi et al. 2003; Yoshida et al. 2004); quasars powered by accretion onto black holes (e.g. Ricotti \& Ostiker 2004; Alvarez \& Abel 2007; Volonteri \& Gnedin 2009); early populations of high-mass X-ray binaries (e.g. Power et al. 2009; Mirabel et al. 2011); high-velocity structure formation shocks (Miniati et al. 2004; Dopita et al. 2011); and dark matter decay and annihilation (e.g. Schleicher et al. 2008).

One additional source of hydrogen ionizations which inevitably accompanies stellar sources are the supernovae (SNe) which mark the end of the life of massive stars.  The impact that SNe have 
on reionization is at least two-fold: not only are ionizing photons generated in SNe explosions (e.g. Chevalier 1974; Shull \& Silk 1979), but the collisional ionization and evacuation of gas they cause lower the optical depth to ionizing radiation, thereby allowing for a higher escape fraction of ionizing photons into the IGM (e.g. Clarke \& Oey 2002; Yajima et al. 2009; Paardekooper et al. 2011; see also Tegmark et al. 1993).  At the high redshifts ($z$ $\sim$ 20) at which the first stars formed, due to the high energy density of CMB photons (which is $\propto$ (1+$z$)$^4$) X-ray photons produced via Compton scattering of the CMB inside SNe remnants are likely to have contributed to the reionizization of the IGM, as described by Oh (2001).  However, over the full redshift range over which reionization takes place (i.e. $z$ $\ga$ 6) SNe generate ionizing photons predominantly through other processes, such as bremsstrahlung and resonance and recombination line emission (e.g. Cox 1972; Chevalier 1974; Shull \& McKee 1979); indeed, these processes can also be dominant even for some Pop III SNe in minihaloes at $z$ $\sim$ 20 (see Kitayama \& Yoshida 2005; Greif et al. 2007; Whalen et al. 2008).   
While, in any case, the number of ionizing photons produced in SNe shocks is in general well below the number produced during the pre-SN phases of stellar evolution, because they are on average more energetic than the photons emitted from stars, the cross section for their absorption by atomic hydrogen and helium is smaller; this, in turn, allows for a higher fraction of SN shock-generated photons to escape into the IGM and contribute to reionization.  

In the present work, we calculate the magnitude of this effect and derive an estimate of the contribution, relative to stellar sources, 
that ionizing photons generated by SNe make to cosmic reionization.  In the next Section, we estimate the number of ionizing photons generated in SNe shocks using recently published high velocity shock models.
In Section 3, we calculate the average cross section for their absorption by neutral hydrogen and helium, and we use this to estimate 
the escape fraction of SNe shock-generated ionizing photons as a function of the escape fractions of stellar photons found in the detailed numerical
simulations of Gnedin et al. (2008) and  Razoumov \& Sommer-Larsen (2010).  In turn, we draw on these results to estimate the relative contributions of stars and SNe to reionization
in Section 4. Though of less overall importance, in Section 5 we furthermore provide an upper limit for the relative contribution of 
cosmic rays generated in SNe shocks to reionization.  In Section 6 we briefly discuss the implications of our findings for the population of galaxies
responsible for reionization.  Finally, we discuss our results and give our concluding remarks in Section 7.

\section{The rate of hydrogen ionization}
We begin by estimating the rates at which hydrogen atoms are ionized due to both SNe and stars, as a function of the star formation rate.  
In the following Sections, we calculate the average cross section for the absorption of these photons by neutral hydrogen and helium, and we 
use these quantities to compare the relative contributions of SN-generated photons and stellar photons to the reionization of the IGM. 

\subsection{Ionization Rate Due to Supernovae}
To estimate the rate at which hydrogen atoms are ionized due to the ionizing radiation emitted from SN remnants, 
we first estimate the total energy emitted in ionizing photons from SN shocks during the early Sedov-Taylor phase of their evolution.  
We then estimate the average energy of the SN shock-generated ionizing photons, and use this to estimate the rate at which 
atomic hydrogen is ionized, as a function of the star formation rate.  

\subsubsection{The Energy Emitted in Ionizing Radiation}
In order to estimate the rate at which ionizing photons are produced in SN shocks we make use of the MAPPINGS III high 
velocity shock models presented in Allen et al. (2008).  We use an approximation of the convenient formula (their equation 5) provided by these authors to 
calculate the flux $F_{\rm ion}$ of hydrogen-ionizing radiation, mostly composed of thermal bremsstrahlung and recombination and resonance lines, 
that is emitted from shocks with velocities 10$^2$ km s$^{-1}$ $\le$ $v_{\rm sh}$ $\le$ 10$^3$ km s$^{-1}$ passing
through a medium with a particle number density 0.01 cm$^{-3}$ $\le$ $n$ $\le$ 10$^2$ cm$^{-3}$:

\begin{equation}
F_{\rm ion} \simeq 2.44 \times 10^{-4} \left(\frac{v_{\rm sh}}{100 \,{\rm km \:s^{-1}}} \right)^3 \left(\frac{n}{1 \, {\rm cm}^{-3}} \right)  {\rm erg \: cm^{-2} \: s^{-1}} \mbox{\ .}
\end{equation}
Next, we make the assumption, justified below in Section 2.1.2, that ionizing photons are produced by SNe shocks while in the Sedov-Taylor phase of their expansion (see e.g. Truelove \& McKee 1999).  
Using the Sedov-Taylor solution, we thus find the following expression for the the distance $r_{\rm sh}$ that the shock has traveled since the initial SN blast, as a function of its velocity $v_{\rm sh}$:

\begin{equation}
r_{\rm sh} = 35  \left(\frac{E_{\rm SN}}{10^{51} \: {\rm erg}} \right)^{\frac{1}{3}} \left(\frac{n}{1 {\rm cm}^{-3}} \right)^{-\frac{1}{3}} \left(\frac{v_{\rm sh}}{100 \,{\rm km \:s^{-1}}} \right)^{-\frac{2}{3}} \, {\rm pc} \mbox{\ .}
\end{equation}
Using equations (1) and (2), we integrate over the whole of the spherical blast wave and over time, to find the total energy $E_{\rm ion}$ emitted in ionizing radiation :

\begin{eqnarray}
E_{\rm ion} & \simeq & \int_{}^{} 4 \pi r^2 F_{\rm ion} dt \nonumber \\
& = & 2.4 \times 10^{50} \, {\rm erg} \left(\frac{E_{\rm SN}}{10^{51} \, {\rm erg}} \right) \int_{200 {\rm km s^{-1}}}^{10^3 {\rm km s^{-1}}} \frac{dv_{\rm sh}}{v_{\rm sh}}        \nonumber \\
& = & 3.8 \times 10^{50} \left(\frac{E_{\rm SN}}{10^{51} \, {\rm erg}} \right)  \, {\rm erg} \mbox{\ ,}
\end{eqnarray}
where in the second part of the equation we have expressed the integral in terms of $v_{\rm sh}$ using equation (2).  
The lower limit of integration is set by the approximate minimum shock velocity ($\sim$ 200 km s$^{-1}$) for which ionizing radiation may propagate beyond the shock, 
assuming a largely neutral medium ahead of the shock (see e.g. Shull \& McKee 1979; Allen et al. 2008); for lower velocities, the ionizing photons produced in the shock
are all absorbed by neutral gas as it is swept up by the shock.  If the ambient medium is instead largely ionized, then ionizing radiation produced in somewhat lower velocity shocks
will propagate beyond the shock as well; thus, our assumption of a neutral ambient medium is a conservative one (see e.g. Shull \& Silk 1979).  
The upper limit of integration in equation (3) is set by the maximum velocity for which the shock models of Allen et al. (2008) 
are given; however, for this reason, too, our estimate of the energy emitted in ionizing radiation is conservative.

Also, interestingly, there is no dependence on the density $n$ of the ambient gas in equation (3).  Rather, it is simply that $\sim$ 40 percent of the energy of the SN is emitted as ionizing radiation, which is 
roughly consistent with previous estimates (see e.g. Mansfield \& Salpeter 1974; Chevalier 1977), as well as being consistent with the amount of energy that is radiated during the Sedov-Taylor phase (see e.g.
Draine \& Woods 1991; Shu 1992), the extent of which we shall turn to discuss in the next Section. 

Before moving on, we must note that the shock models presented in Allen et al. (2008) are, strictly speaking, valid in the steady-state approximation, which may be a poor description of the 
early stages of the Sedov-Taylor phase.  Nonetheless, we find agreement with previous works in part because the majority of both the total energy radiated in ionizing photons and the total number of ionizing photons (see Section 2.1.3) are emitted at the late stages of the Sedov-Taylor phase, when the steady state approximation is more robust.\footnote{The reason that most energy is radiated at the late stages is two-fold: firstly, due to the remnant's slowing expansion with time, the remnant spends more time in the late stages; secondly, the emitting area is much larger at the late stages than at early times.}  Furthermore, we note that even the gas shock-heated to very high temperatures $\ga $ 10$^7$ K at early times will eventually cool radiatively, 
albeit on timescales of up to $\sim$ 10$^6$ yr, considerably longer than the extent of the Sedov-Taylor phase (see e.g. Cioffi et al. 1988).  
Therefore, while it may take upwards of a million years, the post-shock gas will cool and the time-integrated spectrum of 
the radiation emitted from the SNe remnants should converge, at least approximately, 
to the result that we have found under the assumption of steady-state (radiative) shocks.  
As this timescale is still shorter than the $\sim$ 10$^7$ yr timescale on which most massive stars emit ionizing photons and explode as SNe, 
our use of steady-state shock models should not greatly impact our results in the end.

\subsubsection{The Extent of the Sedov-Taylor Phase}
The radius $r_{\rm rad}$ at which a SN shock transitions from the Sedov-Taylor phase to the radiative phase depends on the rate at which the shocked gas cools 
(see e.g. Shull \& Silk 1979; Blondin et al. 1998).  We shall draw on the results of Draine \& Woods (1991), who model the cooling of SN remnants. These authors find the following 
fitting formula for $r_{\rm rad}$, for the case of a gas with solar metallicity (see their tables 1 \& 2):

\begin{equation}
r_{\rm rad} \simeq 32  \left(\frac{E_{\rm SN}}{10^{51} {\rm erg}} \right)^{0.1} \left(\frac{n}{1 \, {\rm cm^{-3}}} \right)^{-0.45} \, {\rm pc} \mbox{\ ,}
\end{equation}
For gas with a metallicity different than solar, the time $t_{\rm rad}$ at which the transition to the radiative phase occurs can be estimated as the cooling time of the shocked gas, and is thus 
inversely proportional to the cooling rate $\Lambda$ of the gas.
In turn, this affects the radius of the transition, as in the energy conserving phase $r_{\rm rad}$ $\propto$ $t_{\rm rad}$$^{2/5}$ $\propto$ $\Lambda$$^{-2/5}$ (see e.g. Blondin et al. 1998). 
To express $r_{\rm rad}$ as a function of metallicity, we can thus follow Blondin et al. (1998), and estimate the cooling function of the gas in the energy conserving phase, as a 
function of metallicity $Z$, by fitting to the cooling functions presented in Sutherland \& Dopita (1993).  We thus find an approximate fit to the cooling function as

\begin{equation}
\Lambda \sim 10^{-22}  \left(\frac{Z}{Z_{\odot}} \right)^{0.44} \left(\frac{T}{10^6 {\rm K}}\right)^{-1} \: {\rm erg \: cm^{3} \: s^{-1}} \mbox{\ ,}
\end{equation}
where $T$ is the temperature of the gas.  This fit is valid over the temperature range $\sim$ 10$^5$ K $\la$ $T$ $\la$ 10$^6$ K, in which the transition to the radiative phase takes place for 
$n$ $\la$ 10$^2$ cm$^{-3}$ (e.g. Draine \& Woods 1991), and over the metallicity range 0.05 $Z_{\odot}$ $\la$ $Z$ $\la$ $Z_{\odot}$, 
which is the same as that over which we model the stellar ionizing photon output below in Section 2.2.  
Using the scaling of $\Lambda$ with metallicity,  we follow 
equations (4.15) and (4.16) of Draine \& Woods (1991) to find the following expression for $r_{\rm rad}$ as a function of metallicity (for $n$ $\la$ 10$^2$ cm$^{-3}$):

\begin{equation}
r_{\rm rad} \simeq 32  \left(\frac{E_{\rm SN}}{10^{51} \, {\rm erg}} \right)^{0.1}  \left(\frac{n}{1 \, {\rm cm^{-3}}} \right)^{-0.45} \left(\frac{Z}{Z_{\odot}} \right)^{-0.05} \, {\rm pc} \mbox{\ .}
\end{equation}
Thus, there is a very weak dependence of the transition radius on metallicity, and a somewhat stronger dependence on the 
density of the gas into which the SN explodes.  Following equation (2), in the Sedov-Taylor phase the distance $r_{\rm sh}$ to which the
shock has propagated, as a function of shock velocity $v_{\rm sh}$, is

\begin{equation}
r_{\rm sh} = 22  \left(\frac{E_{\rm SN}}{10^{51} \, {\rm erg}} \right)^{\frac{1}{3}} \left(\frac{n}{1 \, {\rm cm}^{-3}} \right)^{-\frac{1}{3}} \left(\frac{v_{\rm sh}}{200 \,{\rm km \: s^{-1}}} \right)^{-\frac{2}{3}} \, {\rm pc} \mbox{\ .}
\end{equation}
Thus, the radius at which the velocity drops to 200 km s$^{-1}$, the minimum shock velocity for which we assume ionizing photons can escape the shock front (see Section 2.1.1), is in general smaller than 
the radius at which the Sedov-Taylor phase ends; therefore, our assumption in the previous Section that the Sedov-Taylor phase lasts until $v_{\rm sh}$ = 200 km s$^{-1}$ is sound in most cases.  We note, however,
that for $E_{\rm SN}$ $\ga$ 8 $\times$ 10$^{51}$ erg and/or $n$ $\ga$ 40 cm$^{-3}$, $r_{\rm sh}$ when $v_{\rm sh}$ $\simeq$ 200 km s$^{-1}$ will exceed $r_{\rm rad}$ and our simple
treatment of the dynamical evolution of the SN remnant will have to be modified to account for ionizing photon production after the Sedov-Taylor phase.

We would like to emphasize that, while the Sedov-Taylor solution does well to describe the {\it dynamical} evolution of a SN remnant at early times under the assumption
of conservation of mechanical energy, it is in fact the case that a portion of this energy is radiated away during this phase.  Indeed, the Sedov-Taylor phase ends only when 
a substantial portion of the energy of the SN has been radiated 
away, with the result that the conservation of mechanical energy ceases to be an appropriate approximation for describing the dynamical evolution of the remnant, 
as has been found in myriad studies (see e.g. Chevalier 1974; Draine \& Woods 1991; Blondin et al. 1998; Truelove \& McKee 1999; Shu 2002).  As noted in Section 2.1.1, 
the amount of energy that we have found to be lost to ionizing radiation, especially in the late stages of the Sedov-Taylor phase, is in good agreement with these previous works.

\subsubsection{The Number of Ionizations}
We estimate the total number of ionizing photons emitted per SN using the tabulated ionization parameters given in Table 3 of Allen et al. (2008).  Carrying out an integration 
similar to that shown above, approximated as a sum over shock velocities of $v_{\rm sh}$ = 200 to 10$^3$ km s$^{-1}$, we estimate the total 
number $N_{\rm ion}$ of ionizing photons emitted per SN to be

\begin{equation}
N_{\rm ion} \simeq 4.4 \times 10^{60} \left(\frac{E_{\rm SN}}{10^{51} \, {\rm erg}} \right) \mbox{\ .}
\end{equation}
Dividing $E_{\rm ion}$ by $N_{\rm ion}$ we find the average energy of ionizing photons to be $<E_{\gamma}>$ = 51 eV,
which implies that the majority of the energy of the ionizing radiation is carried by photons with energies $\ga$ 40 eV.
As shown by Shull \& van Steenberg (1985), ionizing photons with such high energies may ionize more than one hydrogen atom per photoionization,  
due to secondary ionizations caused by collisions with photoelectrons.  In particular, these authors show that the fraction of the energy of such ionizing photons that go into 
ionizing hydrogen is $f_{\rm ion}$ $\sim$ 0.3.  We thus estimate the total number of hydrogen atoms that can be ionized by the radiation emitted from a single SN by 
taking it that 30 percent of the energy in ionizing radiation $E_{\rm ion}$ goes into ionizating hydrogen atoms, each of which requires $E_{\rm H}$ = 13.6 eV.  From this, we
find that the total number of ionizations per SN is $f_{\rm ion}$$E_{\rm ion}$/$E_{\rm H}$ $\simeq$ 5.2 $\times$ 10$^{60}$($E_{\rm SN}$/10$^{51}$ erg).  

In turn, we can express the rate $Q_{\rm SN}$ at which hydrogen ionizations take place for a given star formation rate $\dot{M_{\rm *}}$ and a 
stellar initial mass function (IMF) for which $N_{\rm SN}$ supernovae are produced per unit mass in stars, as

\begin{eqnarray}
Q_{\rm SN} & \simeq & 2 \times 10^{51} \left(\frac{\dot{M_{\rm *}}}{{\rm M}_{\odot} \: {\rm yr}^{-1}} \right)\nonumber \\
& \times &   \left(\frac{N_{\rm SN}}{10^{-2} {\rm M}_{\odot}^{-1}} \right) \left(\frac{E_{\rm SN}}{10^{51} \, {\rm erg}} \right) \, {\rm s}^{-1}    \mbox{\ .} 
\end{eqnarray}
We have normalized to $N_{\rm SN}$ = 10$^{-2}$ M$_{\odot}^{-1}$, which is approximately the value for a standard Salpeter-like IMF (e.g. Leitherer et al. 1999).   
We note that equation (9) takes into account only type II SNe produced by stars with masses $\ga$ 9 M$_{\odot}$, and not type Ia SNe; this is reasonable, however, 
because the number of type Ia SNe, which typically occur with a significant delay following a star formation episode, is in general much smaller than that of type II SNe (see e.g. Nagashima et al. 2005).

\subsection{Ionization Rate Due to Stars}
Next, we estimate the rate $Q_{\rm *}$ at which ionizing photons are produced by stars, as a function of $\dot{M_{\rm *}}$, assuming 
a given number of ionizing photons $\eta$ emitted per baryon in stars.  Normalizing $\eta$ to a value appropriate for Pop~II stars formed with a 
standard Salpeter-like IMF (e.g. Haiman 2009), we have 

\begin{equation}
Q_{\rm *} \simeq 10^{53}  \left(\frac{\dot{M_{\rm *}}}{{\rm M}_{\odot} {\rm yr}^{-1}} \right) \left(\frac{\eta}{4 \times 10^{3}} \right) \, {\rm s}^{-1} \mbox{\ .}
\end{equation}
The quantity $\eta$ is metallicity and IMF dependent, and can, for instance, be upwards of 10 times higher than the fiducial value we have chosen here for
a population of extremely low- or zero-metallicity stars (e.g. Bromm et al. 2001; Schaerer 2002; Wise \& Cen 2009).
In practice, we shall use the values of $Q_{\rm *}$ given for a range of metallicities in the Starburst99 database (Leitherer et al. 1999);
specifically, we shall use the models from these authors for stellar populations formed with a Salpeter IMF, with minimum and maximum initial stellar masses of 1 and 100 M$_{\odot}$.  
These choices will allow to facilitate a direct comparison with radiative transfer simulations from which the escape fractions of
ionizing photons are calculated, as described in the next Section.  

Comparing $Q_{\rm SN}$ to $Q_{\rm *}$, we see that the rate at which hydrogen atoms are ionized due to radiation from stars is 
in general more than an order of magnitude higher than that due to the emission of ionizing radiation from SN shocks.  Nonetheless, the latter
can contribute significantly to reionization due to a higher fraction of these photons escaping from galaxies, which we now turn to show.

\section{The Escape Fraction of SNe Shock-generated Ionizing Photons}
In this Section we first calculate the average photoionization cross section for SNe shock-generated photons, and we then
apply this to attain an estimate of the escape fraction of such photons as compared to that of stellar ionizing photons.

\subsection{The Average Cross Section for Photoionization}

In order to estimate the average cross section for photoionization of SN shock-generated photons, we make use of the high velocity shock spectra presented in Allen et al. (2008).  As shown
in Section 2.1.3, the average energy of ionizing photons from SNe shocks is high enough that we need to account for multiple hydrogen ionizations per photon, as we have done in calculating $Q_{\rm SN}$.  
As the number of ionizations is thus dependent on the total energy in ionizing radiation, it is the fraction of the energy in ionizing photons which escapes the host galaxy which
determines the impact of SN shock-generated radiation on the reionization of the IGM. 
Therefore, we estimate the cross section $\sigma_{\rm SN}$ for photoionization as the {\it photon energy}-averaged cross section:

\begin{equation}
\sigma_{\rm SN} \simeq \frac{\int_{{\rm 13.6 eV}}^{\infty} \left[\sigma_{\rm HI}(E_{\gamma}) + \sigma_{\rm HeI}(E_{\gamma}) \right] F(E_{\gamma}) dE_{\gamma}}{\int_{{\rm 13.6 eV}}^{\infty} F(E_{\gamma}) dE_{\gamma}} \mbox{\ ,}
\end{equation}
where the spectrum is described by $F$($E_{\gamma}$), the flux per unit photon energy.  
We use the photoionization absorption cross sections for neutral hydrogen and helium provided by Osterbrock \& Ferland (2006). 
In particular, for photon energies $E_{\gamma}$ $\ge$ 13.6 eV we take the cross section 
for absorption by neutral hydrogen to be $\sigma_{\rm HI}$ = 6.3 $\times$ 10$^{-18}$ cm$^{2}$ ($E_{\gamma}$/13.6 eV)$^{-3}$, whereas for photon energies above 24.6 eV we compute the total cross section for absorption as the sum of the neutral hydrogen and neutral helium cross sections for absorption, the latter taken to be $\sigma_{\rm HeI}$ = 7.8 $\times$ 10$^{-18}$ cm$^{2}$ ($E_{\gamma}$/24.6 eV)$^{-3}$. We have assumed a neutral helium to hydrogen ratio of 0.1, such that our results
are obtained as the average cross section to absorption per neutral hydrogen atom, which is a convenient form for the calculation 
of the escape fraction of SN shock-generated photons which follows.  We note that this assumption implies a negligible amount of absorption of photons by helium which is already once-ionized 
(i.e. in the form of He~{\sc ii}).  Given the relatively soft spectra of the stellar sources, the fraction of helium in the form of He~{\sc ii} is likely to be low;  
given furthermore the low photoionization cross section of He~{\sc ii} compared to that of He~{\sc i} (e.g. Osterbrock \& Ferland 2006), this is a conservative approximation.

The average cross section for photoionization that we obtain following this procedure is $\sigma_{\rm SN}$ $\sim$ 0.7 $\times$ 10$^{-18}$ cm$^{2}$, for the solar metallicity shock spectra 
presented in Allen et al. (2008).  As these authors note, the strength of the ionizing field, being largely generated by brehmsstrahlung, is only weakly dependent on the metallicity of the gas; therefore, 
we shall use this single value for our calculations for simplicity, although we will express our results in terms of $\sigma_{\rm SN}$ where possible, such that
it is clear how our results will change for different average cross sections.

Next, in order to estimate the escape fraction of SN shock-generated photons relative to that of stellar photons, 
we also need to calculate the average cross section $\sigma_{\rm *}$ for photoionization of stellar ionizing photons.  
For this we adopt the model spectra presented in Leitherer et al. (1999).  In particular, we use the spectra for continuous star formation
at metallicities of $Z$ = 0.05, 0.2, 0.4, and 1 $Z_{\odot}$, and a Salpeter IMF over a mass range from 1 to 100 M$_{\odot}$.

We carry out the same exercise as described above for the calculation of $\sigma_{\rm SN}$, but we now instead calculate the {\it photon number}-averaged cross section for photoionization.
This is appropriate in this case, because for the softer spectra of ionizing radiation emitted from stars a single ionizing photon will on average lead to the ionization of a single hydrogen atom.
We thus calculate the cross section as 

\begin{equation}
\sigma_{\rm *} \simeq \frac{\int_{{\rm 13.6 eV}}^{\infty} \left[\sigma_{\rm HI}(E_{\gamma}) + \sigma_{\rm HeI}(E_{\gamma}) \right] \frac{F(E_{\gamma})}{E_{\gamma}} dE_{\gamma}}{\int_{{\rm 13.6 eV}}^{\infty} \frac{F(E_{\gamma})}{E_{\gamma}} dE_{\gamma}} \mbox{\ ,}
\end{equation}
For each of the stellar spectra we consider, the values we find for $\sigma_{\rm *}$ range from 2.6 $\times$ 10$^{-18}$ cm$^2$ to 4.8 $\times$ 10$^{-18}$ cm$^2$, as shown in Table 1.  
Due to the softer ionizing spectra of stellar populations at higher metallicity, for higher values of the metallicity the average cross section is found to be higher as well.  
Of particular use to us will be the average cross section for a stellar metallicity of $Z$ = 0.05 $Z_{\odot}$, as this is the same population of stars used in the radiative 
transfer simulations which we shall draw on in the next Section to estimate the escape fractions of SN shock-generated ionizing photons (Gnedin et al. 2008; Razoumov \& Sommer-Larsen 2010; see also Yajima et al. 2011).

\begin{figure*}
  \centering
  \includegraphics[width=5.0in]{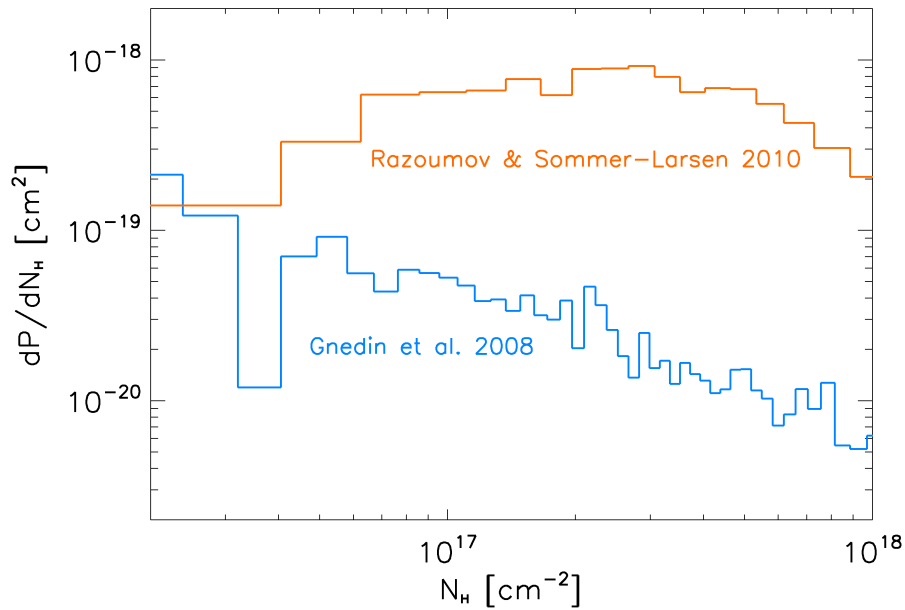}
  \caption
  {Probability distribution of neutral hydrogen column densities $N_{\rm H}$ derived from the radiative transfer simulations presented in Gnedin et al. (2008) ({\it blue line}) and Razoumov \& Sommer-Larsen (2010) ({\it red line}).  The transformation from the escape fraction probability distributions presented by these authors is described in Section 3.2.2.  Note that, as these authors report, there is a large probability that $f_{\rm esc,*}$ = 0 or $f_{\rm esc, *}$ = 1; however, as these values imply $N_{\rm H}$ = $\infty$ and $N_{\rm H}$ = 0, respectively, these points are not shown in this logarithmic plot.  Nonetheless, they are accounted for 
in the calculations of the escape fractions of stellar and SN shock-generated ionizing photons presented in Section 3.2.2.}
\end{figure*}

\subsection{The Escape Fractions of Stellar and SNe Photons}
In order to compare the escape fractions of stellar and SN shock-generated ionizing photons, we first estimate
their ratio along a given line of sight out of a galaxy into the IGM.  We then use this result, in conjunction with 
the results of detailed radiative transfer simulations, to estimate the overall, line of sight-averaged escape fractions of ionizing 
photons.

\subsubsection{Escape Fractions Along a Single Line of Sight}
To begin, we note that the escape fraction of ionizing photons along a given line of sight out of a galaxy can be approximately 
expressed as\footnote{In the appendix, we show explicitly the approximation we have made in order to arrive at this formula.  The reason 
that we do not use the exact formula for the escape fraction, which is the integral over $E_{\rm \gamma}$ of e$^{-N_{\rm H}\sigma(E_{\rm \gamma})}$ weighted by the source
spectrum, is that we must extract the probability distribution function of the hydrogen column density d$P$/d$N_{\rm H}$ from the escape 
fractions reported in Gnedin et al. (2008) and Razoumov \& Sommer-Larsen (2010), in order to obtain the relative escape fraction 
of SNe ionizing photons.  This is made possible only by using the approximate formula given by equation (13).}

\begin{equation}
f_{\rm esc} = e^{-\tau} = e^{-N_{\rm H}\sigma}  \mbox{\ ,}
\end{equation}
where $\tau$ is the optical depth along the line of sight, which is given in terms of the column density $N_{\rm H}$ of neutral hydrogen 
and the average cross section $\sigma$ for the absorption of ionizing photons by neutral hydrogen and helium.  Assuming that the location of 
a population of  stars in the galaxy is the same as that of the SNe that mark their deaths (i.e. that the spatial extent 
of radiating SNe remnants is sufficiently small) and that the column density $N_{\rm H}$ along a given line of sight out of the galaxy
does not change significantly between the time when a population of stars is shining and when they explode as 
SN (i.e. within $\la$ 3 $\times$ 10$^7$ yr; e.g. Leitherer et al. 1999), 
the column density will be the roughly same for both cases.  Assuming that it is in fact the same, we have for the ratio of the escape fractions of SNe and stellar ionizing photons

\begin{equation}
\frac{f_{\rm esc, SN}}{f_{\rm esc, *}} \simeq \frac{e^{-\tau_{\rm SN}}}{e^{-\tau_{\rm *}}} = e^{\tau_{\rm *}(1-\frac{\sigma_{\rm SN}}{\sigma_{\rm *}})} \mbox{\ ,}
\end{equation}
where $\tau_{\rm *}$ = $N_{\rm H}$$\sigma_{\rm *}$ and $\tau_{\rm SN}$ = $N_{\rm H}$$\sigma_{\rm SN}$.
In turn, using once again $f_{\rm esc, *}$ = e$^{-\tau_{\rm *}}$, we solve for the escape fraction of SN photons:

\begin{equation}
f_{\rm esc, SN} \simeq e^{-\tau_{\rm *}\frac{\sigma_{\rm SN}}{\sigma_{\rm *}}} = f_{\rm esc, *}^{\frac{\sigma_{\rm SN}}{\sigma_{\rm *}}} \simeq f_{\rm esc, *}^{0.27} \mbox{\ ,}
\end{equation}
where for the final expression we have used the values of $\sigma_{\rm SN}$ and $\sigma_{\rm *}$ found in Section 3.1 for the case of star formation
at a constant rate with a Salpeter mass function from 1 - 100 M$_{\odot}$ and a metallicity of $Z$ = 0.05 $Z_{\odot}$, as presented in Leitherer et al. (1999).

We can see immediately from equation (15) that, especially for low escape fractions of stellar photons along a given line of sight, the escape fraction of SNe photons
can be much larger than that of stellar photons.  For example, using the last expression in equation (15), 
for $f_{\rm esc, *}$ = 0.1 we obtain $f_{\rm esc, SN}$ = 0.54, and for $f_{\rm esc, *}$ = 0.01 we obtain $f_{\rm esc, SN}$ = 0.29.
Taking the ratio of the number of ionizations caused by SNe and stars found in Section 2 ($Q_{\rm SN}$/$Q_{\rm *}$ $\sim$ 0.02), we see that more reionizations due to SNe than to stellar photons 
will take place in the IGM (i.e. that $f_{\rm esc, *}$ $Q_{\rm *}$ $\la$ $f_{\rm esc, SN}$ $Q_{\rm SN}$)
when the escape fraction of stellar photons is $f_{\rm esc, *}$ $\la$ 0.005; this corresponds to a hydrogen column density along the line of sight out of a galaxy of  $N_{\rm H}$ $\simeq$ 10$^{18}$ cm$^{-2}$.
For column densities greater than this, the contribution to reionization from SNe is greater than that from stars.

While in deriving equation (15) we have assumed both a constant star formation rate and no evolution in the column density along a given line of sight over a 
timescale of $\la$ 3 $\times$ 10$^7$ yr, roughly the time between the emission of the bulk of the ionizing photons from stars and the emission from SNe remnants, 
in general these assumptions may lead us to underestimate the escape fraction of SNe-generated photons.  The reason for this can be seen by considering a single stellar 
population formed instantaneously.  In this case, the high energy photons emitted by stars will act to heat and ionize the gas 
along a line of sight out of the galaxy, and the effect of this stellar feedback will be to increase the escape fraction of ionizing photons with time (see e.g. Johnson et al. 2009; Wise \& Cen 2009).
Because this feedback occurs {\it before} the ionizing photons from SNe are emitted, it is likely that in general the column density along a line of sight out of a galaxy
from a given cluster of stars is lower when SNe occur than when the bulk of the ionizing photons from stars are emitted.  Therefore, it is likely the case 
that the escape fractions of SN-generated photons are larger relative to the time-averaged stellar photon escape fractions than we have found here.

\subsubsection{Line of Sight-Averaged Escape Fractions}
To better estimate the relative numbers of hydrogen ionizations caused by SNe and stars, we turn to the results of detailed radiative transfer simulations from 
which the escape fraction of stellar photons has been calculated.  As our result for the ratio of escape fractions is defined only along a given line of sight, we 
must use the results of numerical simulations which are likewise given along a representative collection of lines of sight.  To this end, we use the probability distributions for 
the escape fractions of stellar ionizing photons given in figure 2 of Gnedin et al. (2008) and figure 9 of Razoumov \& Sommer-Larsen (2010)\footnote{We note that probability distribution functions similar to those of Razoumov \& Sommer-Larsen (2010) have recently been presented in Yajima et al. (2011).}.

As we would like to generalize our results
for the various values of $\sigma_{\rm *}$ given in Table 1, we shall convert the probability distributions of escape fractions which are given by these authors to the 
probability distributions of neutral hydrogen column densities $N_{\rm H}$.  To do this, we first note that from equation (13) we have $N_{\rm H}$ = -ln($f_{\rm esc, *}$)/$\sigma_{\rm *}$,
where $f_{\rm esc, *}$ is the stellar photon escape fraction derived from the simulations and $\sigma_{\rm *}$ is the average cross section for photoionization that we have calculated for the 
same stellar spectra used in these simulations, the $Z$ = 0.05 $Z_{\odot}$ model from Leitherer et al. (1999).  Using this, we next convert from the probability distribution $dP$/$df_{\rm esc, *}$
of stellar ionizing photons to the probability distribution of neutral hydrogen column densities $dP$/$dN_{\rm H}$, as follows:

\begin{equation}
\frac{dP}{dN_{\rm H}} = \sigma_{\rm *} f_{\rm esc, *} \left| \frac{dP}{df_{\rm esc,*}} \right| \mbox{\ .}
\end{equation}
The resulting probability distributions $dP$/$dN_{\rm H}$ that we find for the simulations presented in Gnedin et al. (2008) and Razoumov \& Sommer-Larsen (2010) are shown in Figure 1.
These probability distributions contain the basic information that we would like to glean from these detailed hydrodynamics simulations. Using these, we can next apply different cross sections $\sigma_{\rm *}$ 
and $\sigma_{\rm SN}$ to infer the probability distributions of the escape fractions of stellar and SN shock-generated ionizing photons, and from this the overall escape fractions averaged over all lines of sight.
We note that in our calculation of $N_{\rm H}$ we have neglected the absorption of ionizing photons by dust; however, as Gnedin et al. (2008) show, this is a relatively minor effect and hence 
should not impact our results greatly (see also Yajima et al. 2011).

\begin{figure*}
  \centering
  \includegraphics[width=5.0in]{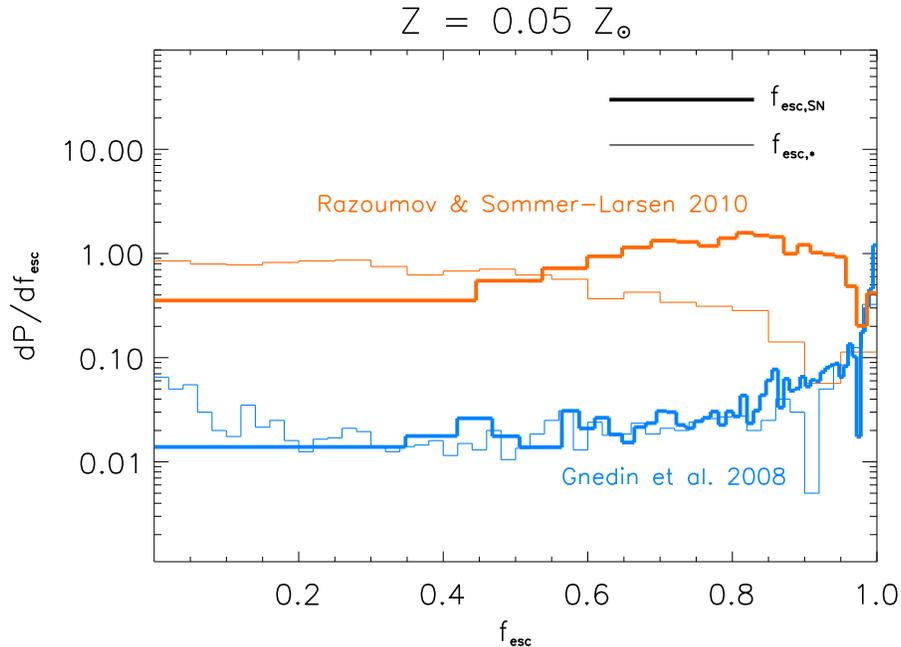}
  \caption
  {Probability distribution functions of escape fractions $f_{\rm esc}$ of ionizing photons from stars ({\it thin lines}) and from SNe remnants ({\it thick lines}), assuming stellar spectra from a population of stars at metallicity $Z$ = 0.05 $Z_{\odot}$, formed with an 1 - 100 M$_{\odot}$ Salpeter IMF (Leitherer et al. 1999). The distribution functions of escape fractions $f_{\rm esc,*}$ of ionizing photons emitted from stars are taken from the results of radiative transfer simulations presented in Gnedin et al. (2008) ({\it blue lines}) and Razoumov \& Sommer-Larsen (2010) ({\it red lines}).  The transformation of the probability distributions of $f_{\rm esc, *}$ to those of SNe-generated ionizing photons $f_{\rm esc, SN}$ is described in Section 3.2.2.  The overall escape fractions, averaged over all lines of sight, of SNe and stellar photons are 0.076 and 0.021, respectively, for the results of Gnedin et al. (2008); those for the results of Razoumov \& Sommer-Larsen (2010) these are 0.80 and 0.22, respectively.  The factor of $\sim$ 4 difference between these average values of $f_{\rm esc,*}$ and $f_{\rm esc,SN}$ is due to the higher average energy, and so lower cross section for absorption, of the photons generated in SNe shocks.  Here, this effect is evident in the probability distributions of SNe photons being shifted to higher escape fractions relative to those of stellar photons. 
}
\end{figure*}

\begin{figure*}
  \centering
  \includegraphics[width=5.0in]{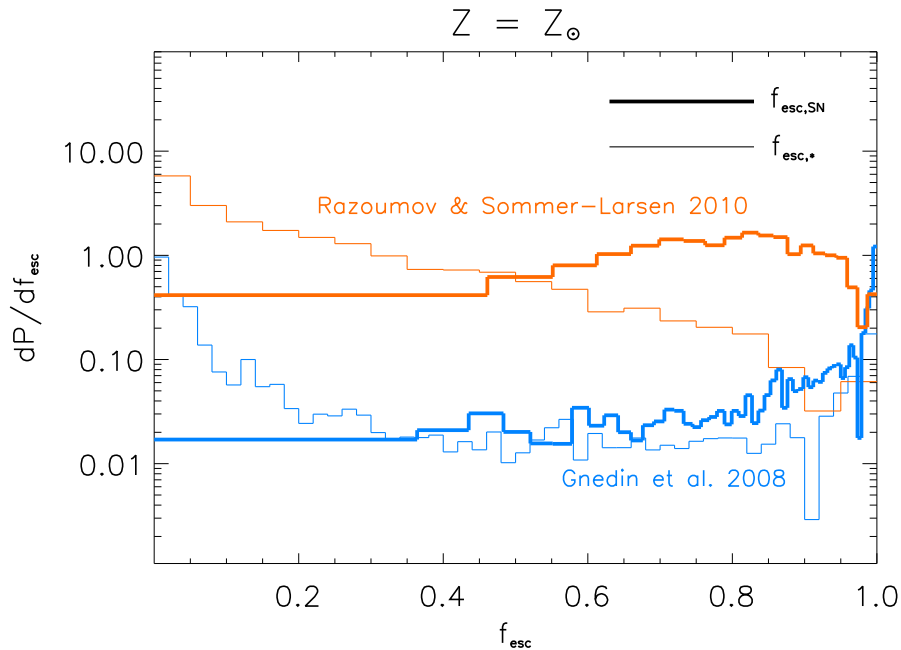}
  \caption
  {The same as Figure 2, but now assuming stellar spectra from a population of stars at metallicity $Z$ = $Z_{\odot}$, formed with an 1 - 100 M$_{\odot}$ Salpeter IMF (Leitherer et al. 1999).   For this case, the overall escape fractions, averaged over all lines of sight, of SNe and stellar photons are 0.076 and 0.011, respectively, for the results of Gnedin et al. (2008); those for the results of Razoumov \& Sommer-Larsen (2010) these are 0.8 and 0.12, respectively.  The factor of $\sim$ 7 difference between the average values for $f_{\rm esc,*}$ and $f_{\rm esc,SN}$ is much larger than in the case of the $Z$ = 0.05 $Z_{\odot}$ spectra shown in Figure 2, due to the lower average energy, and so higher cross section for absorption, of the photons from the higher metallicity stars.}
\end{figure*}

In order to directly compare the probability distributions of the two escape fractions $f_{\rm esc,*}$ and $f_{\rm esc,SN}$, we must convert from that for stellar photons $dP$/$df_{\rm esc,*}$ given
in Gnedin et al. (2008) and Razoumov \& Sommer-Larsen (2010) to that for SNe photons.  For the stellar spectra used in these works, the $Z$ = 0.05 $Z_{\odot}$ stellar population for which we have 
calculated the average cross section to photoionization, this is done as follows:

\begin{equation}
\frac{dP}{df_{\rm esc,SN}} = \frac{df_{\rm esc,*}}{df_{\rm esc,SN}}\left| \frac{dP}{df_{\rm esc,*}} \right| = 3.7 f_{\rm esc, SN}^{2.7} \left| \frac{dP}{df_{\rm esc,*}} \right| \mbox{\ ,}
\end{equation}
where we have used equation (15) relating $f_{\rm esc,SN}$ to $f_{\rm esc,*}$.  With this conversion from the stellar to the SNe photon escape fraction probability distribution, we can 
directly compare the two distributions, which we show together in Figure 2.  As the Figure shows, there is a higher probability for larger escape fractions of SN shock-generated photons than 
for stellar photons.  In Figure 3, we show the probability distributions of $f_{\rm esc, *}$ and $f_{\rm esc, SN}$ for the case of a stellar population at solar metallicity, using the 
appropriate $\sigma_{\rm *}$ from Table 1, along with the corresponding transformation to the probability distribution $dP$/$df_{\rm esc,*}$ for this average cross section.  Comparing the two Figures
it is evident that the probability distributions for $f_{\rm esc,SN}$ and $f_{\rm esc, *}$ are much more different in the case of the solar metallicity stellar population than in the case of the 
lower metallicity stellar population.  This is simply due to the larger difference between the cross sections $\sigma_{\rm *}$ and $\sigma_{\rm SN}$ for the more metal rich population; 
because the spectra of ionizing radiation from more metal-rich stars is softer, the average cross section for photoionization is higher, and so the escape fraction of stellar ionizing photons is
lower relative to the escape fraction of SN shock-generated photons.

\begin{table*} 
\begin{center}
\caption{Relative contributions of stars and Supernovae to reionization}
\begin{tabular}{cccccc}
\tableline
\tableline
& & & & & \\
Metallicity\tablenotemark{a} &  $Q_{\rm *}$\tablenotemark{b} & $N_{\rm SN}$ & $\sigma_{\rm *}$ & Ratio of escape fractions\tablenotemark{c}: & Relative contribution of SNe:\\
$[$$Z_{\odot}$$]$ &  $[$10$^{53}$ s$^{-1}$$]$  & $[$10$^{-2}$ M$_{\odot}^{-1}$$]$ & $[$10$^{-18}$ cm$^{2}$$]$ &  $<f_{\rm esc,SN}>$/$<f_{\rm esc,*}>$ & $<f_{\rm esc,SN}>$$Q_{\rm SN}$/$<f_{\rm esc,*}>$$Q_{\rm *}$\tablenotemark{d} \\ 
& & & & \\
\tableline
& & & & & \\
1 & 2.2  & 1.91 & 4.8 & 6.8 & 0.12\\
0.4 & 2.5  & 1.86 & 2.9 & 4.2 & 0.062\\
0.2 & 2.8  & 1.91 & 2.8 & 4.1 & 0.056\\
0.05 & 3.2  & 1.95 & 2.6 & 4.0 & 0.048\\
& & & & \\
\tableline
\end{tabular}
\tablenotetext{a}{The metallicity of the stellar population.  Here we take the solar metallicity to be $Z_{\odot}$ = 0.02.}
\tablenotetext{b}{The number of ionizing photons per second produced for a constant star formation rate of 1 M$_{\odot}$ yr$^{-1}$.}
\tablenotetext{c}{Estimated from the column density probability distributions derived from the results presented in Gnedin et al. (2008) and Razoumov \& Sommer-Larsen (2010), as described in Section 3.2.2.}
\tablenotetext{d}{Note that the relative contribution of SNe scales linearly with the average SN energy $E_{\rm SN}$; here we assume that $E_{\rm SN}$ = 10$^{51}$ erg.}
\tablecomments{
All stellar populations modeled here are assumed to have a Salpeter IMF, with stars formed in the mass range 1 - 100 M$_{\odot}$.  The values of $Q_{\rm *}$, $N_{\rm SN}$,
 and $\sigma_{\rm *}$ were computed from the models of these stellar populations presented in the Starburst99 database (Leitherer et al. 1999), assuming a constant star formation rate. 
Finally, we have assumed a constant $\sigma_{\rm SN}$ = 0.7 $\times$ 10$^{-18}$ cm$^{2}$, independent of metallicity, as discussed in Section 3.1.
}
\end{center}
\end{table*}

Using our results for the probability distributions of both $f_{\rm esc,*}$ and $f_{\rm esc,SN}$ we calculate the overall escape fractions, averaged over all lines of sight, of SNe and stellar photons.  For the case of SNe photons, this calculation is done as follows:

\begin{equation}<f_{\rm esc,SN}> = \int^{1}_{0} f_{\rm esc,SN} \frac{dP}{df_{\rm esc,SN}} df_{\rm esc, SN} \mbox{\ ,}
\end{equation}
where d$P$/d$f_{\rm esc,SN}$ is given by equation (17).  A similar calculation is carried out for the line of sight-averaged escape fraction of stellar photons $<f_{\rm esc,*}>$, 
with the integral performed over $f_{\rm esc,*}$ instead of $f_{\rm esc,SN}$.

The ratios of the line of sight averaged escape fractions that we find for the stellar populations that we consider 
are similar for both of the simulations that we have drawn upon, and we present these ratios in the fifth column of Table 1.  As shown there,
depending on the metallicity of the stellar population, we find that the values of $f_{\rm esc,SN}$/ $f_{\rm esc,*}$ vary between $\simeq$ 4 and $\simeq$ 7.
We emphasize that, especially for the results of Gnedin et al. (2008), the SNe photon escape fraction is significantly lower than what would be found by simply applying equation (15) to the overall 
stellar escape fraction directly.  This is likely due in part to the fact that the simulation results show relatively large probabilities for escape fractions of either
$f_{\rm esc, *}$ $=$ 1 or $f_{\rm esc, *}$ $=$ 0; because these limiting values will yield the same escape fractions for SN generated photons, via equation (15), the overall line of sight-averaged
escape fractions of stellar and SNe photons are not as different as they are along individual lines of sight for which 0 $<$ $f_{\rm esc,*}$ $<$ 1.

Finally, we note that our results are senstively dependent on the probability distribution functions that we have adopted from Gnedin et al. (2008) and Razoumov \& Sommer-Larsen (2010).  While the detailed simulations which these authors carry out in attaining these results are some of the most sophisticated and complete to date (including radiative feedback from young stars and both mechanical and chemical feedback from SNe), the resolution of the simulations is limited and the structure of the interstellar medium at sub-resolution scales will have an impact on the actual escape fractions of ionizing photons. Despite this uncertainty which plagues all escape fraction calculations, the results presented by these authors allow us to make a first estimate of the relative role of SNe in cosmic reionization.  Future simulations at higher resolution will ideally include the ionizing radiation from both stars and SNe explicitly, thereby allowing for improved estimates of the relative role of each of these sources.

\section{The Relative Contributions of Stars and SNe to reionization}
Working from our results from Sections 2 and 3, we may now compare the relative contributions of stars and SNe to reionization.  The ratio of the number of hydrogen ionizations in the 
IGM due to SNe to that due to stars is given by $<f_{\rm esc, SN}>$$Q_{\rm esc,SN}$/$<f_{\rm esc,*}>$$Q_{\rm esc,*}$.  Using equations (9) and (10), along with the values of $Q_{\rm *}$ and $N_{\rm SN}$ from Leitherer et al. (1999) and 
the ratios of line of sight-averaged escape fractions, each given in Table 1, we estimate this ratio as

\begin{eqnarray}
\frac{<f_{\rm esc,SN}> Q_{\rm SN}}{<f_{\rm esc, *}> Q_{\rm *}} & \simeq & 0.1 \left( \frac{Q_{\rm *}}{10^{53} {\rm s^{-1}}}\right)^{-1}  \nonumber \\
& \times & \left( \frac{N_{\rm SN}}{10^{-2} {\rm M_{\odot}^{-1}}}\right)   \left(\frac{E_{\rm SN}}{10^{51} {\rm erg}}\right)  \nonumber \\
& \times & \left(\frac{<f_{\rm esc,SN}>}{0.4}\right)  \left(\frac{<f_{\rm esc,*}>}{0.1}\right)^{-1} \mbox{\ .}
\end{eqnarray}
The escape fractions appearing here are the overall line of sight-averaged values, normalized to typical values.  In particular, we have normalized the ratio of escape fractions to $<f_{\rm esc,SN}>$/$<f_{\rm esc,*}>$ = 4, roughly the ratio found in the last Section for a stellar population at sub-solar metallicity.  

The right-most column of Table 1 shows the values of this ratio for all of the stellar populations we consider.  The values that we find for the contribution of SNe to reionization, relative 
to the contribution from stars, ranges from $\simeq$ 0.05 to $\simeq$ 0.12, with higher relative contributions for the more metal-rich stellar populations.  As discussed in Section 3, this is due largely
to the higher ratios of escape fractions $<f_{\rm esc, SN}>$/$<f_{\rm esc,*}>$ for the more metal-enriched populations, owing to their softer stellar spectra.  However, the different rates of SN 
production $N_{\rm SN}$ and stellar ionizing photon generation $Q_{\rm *}$ also have an effect.  Importantly, we note that the relative contributions to reionization scale linearly with the energy per supernova $E_{\rm SN}$;
while we have normalized our results to $E_{\rm SN}$ = 10$^{51}$ erg, if SNe are more energetic than this in the early universe, then the relative contribution of SNe to reionization may be 
considerably higher.

 Finally, we note that our calculations of the quantities appearing in Table 1 have been done drawing on numerous approximations.  We have nonetheless chosen to report the values of these quantities, in cases, to two or three decimal places, in order to highlight the differences between the stellar models at different metallicities that we have considered.

\section{The Contribution of Supernovae-generated cosmic rays to reionization}
An additional process driven by SNe which leads to ionization of hydrogen is the production of cosmic rays, and for completeness here we consider their 
contribution to reionization.  We may find an upper limit to the total number of ionizations that cosmic rays can cause in the IGM,
by making the simple assumption that all the energy injected into cosmic rays by SNe is deposited into the IGM via ionizations of hydrogen.  Following the discussion of cosmic ray production
in the early universe given by Stacy \& Bromm (2007), we shall assume that 10 percent of the energy $E_{\rm SN}$ of SNe goes into cosmic rays (e.g. Ruderman 1974); then, noting that each ionization
removes 50 eV of energy from the impinging cosmic ray (Spitzer \& Tomasko 1968), we find the maximum number of ionizations that can be produced per SN to be $\sim$ 10$^{60}$.  
Relating this to the star formation rate, in order to compare the maximum rate (assuming an escape fraction of cosmic rays of $f_{\rm esc,CR}$ = 1) 
of ionizations $Q_{\rm CR}$ by cosmic rays to the rate of ionizations due to stellar and SN shock-generated photons, we have 

\begin{eqnarray}
Q_{\rm CR} & \simeq & 4 \times 10^{50}  \left(\frac{\dot{M_{\rm *}}}{{\rm M}_{\odot} \:{\rm yr}^{-1}} \right)  \nonumber \\
& \times &  \left(\frac{N_{\rm SN}}{10^{-2} \, {\rm M}_{\odot}^{-1}} \right) \left(\frac{E_{\rm SN}}{10^{51} \, {\rm erg}} \right) \, {\rm s}^{-1} \mbox{\ .} 
\end{eqnarray}

We emphasize that this is an extremely hard upper limit, as in general the IGM is optically thin to cosmic rays, especially to those with higher energies.  To illustrate this, 
we can make a simple conversion of the result presented in figure 1 of Stacy \& Bromm (2007) to find the penetration depth $d_{\rm CR}$ of a cosmic ray with energy $\epsilon_{\rm CR}$ traveling 
through a completely neutral general IGM at a redshift $z$:

\begin{equation}
d_{\rm CR} \sim 10  \left(\frac{1+z}{10}\right)^{-3} \left(\frac{\epsilon_{\rm CR}}{10^6 {\rm eV}} \right)^2 \, {\rm Mpc} \mbox{\ ,}
\end{equation} 
where we have applied a simple power-law fit appropriate for low cosmic ray energies (i.e. $\epsilon_{\rm CR}$ $\la$ 10$^8$ eV), and the distance is in physical units.   
Comparing this to the physical distance that a cosmic ray moving at close to the speed of light would travel from its generation at $z$ $\la$ 20 in a SN explosion to the end of reionization at 
$z$ $\sim$ 6, which is $\la$ 200 Mpc, we see that only cosmic rays with $\epsilon_{\rm CR}$ $\la$ 10$^7$ eV will have time to deposit a large portion of their 
energy into ionizing the IGM.  Although the energy distribution of cosmic rays in the early universe is not known, it is likely that the bulk of the total energy in cosmic rays 
generated before reionization is carried by cosmic rays at higher energies that this (e.g. Jasche et al. 2007; Stacy \& Bromm 2007).   Therefore, comparing 
$Q_{\rm CR}$ to $Q_{\rm *}$ and $Q_{\rm SN}$, we conclude that the contribution to reionization from cosmic rays generated in SNe 
is likely to be well below that from either stellar sources or from the radiation generated in SNe shocks.

\begin{figure*}
  \centering
  \includegraphics[width=3.25in]{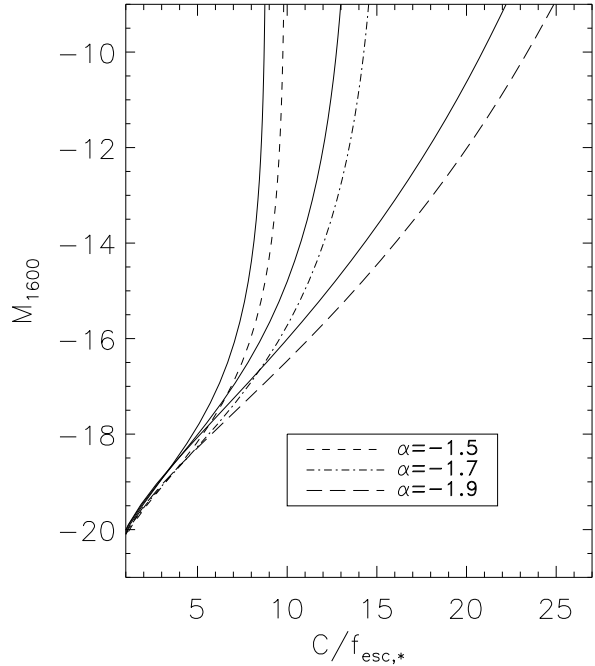}
  \caption
  {The limiting magnitude down to which the luminosity
function needs to be integrated to achieve reionization at $z=7$ for
a given ratio of clumping factor $C$ to escape fraction of stellar photons $f_{\rm esc,*}$. 
Solid lines show results for stellar photons only assuming the fits to the
luminosity function presented in Wilkins et al. (2011). Faint-end slopes
$\alpha$ are increasingly steep, going from left to right. Short-dashed,
dot-dashed and long-dashed lines show the corresponding cases
including the contribution from SN shock-generated photons to reionization. Note that this Figure differs from the similar figure 17 of Wilkins et al. (2011), who used
$D_{\rm ion}=1.47$ instead of $D_{\rm ion}=1.2$ (Wilkins, private communication).}
\end{figure*}

\section{Implications for Reionization by Galaxies}
In the previous Sections we have estimated the relative contributions of
SNe and stars to reionization. We now consider the implications
of our results for the population of galaxies responsible for reionization.  In particular,
we highlight how constraints on the luminosity function of
reionizing galaxies are strengthened by accounting for the contribution of SNe.

Following Wilkins et al. (2011), we assume that the number of ionizing
photons produced by a galaxy of luminosity $L_{1600}$ at  1600 ${\rm \AA}$
 is $Q_{\rm *} = D_{\rm ion} \times 10^{13} L_{1600}$, where the units of $L_{\rm 1600}$ is erg s$^{-1}$ ${\rm \AA}$$^{-1}$. To facilitate a comparison with this work, we use for the
fraction of ionizing photons $D_{\rm ion}=1.2$.  This is appropriate for a
stellar population with metallicity $Z=Z_{\odot}$, with stars having formed with a Salpeter IMF at a constant rate for 100 Myr (Leitherer et al. 1999), as these authors consider. 
The rate of ionizing photon production per unit comoving volume that is required to maintain reionization is given by (Madau et al. 1999; Wilkins 2011):

\begin{equation}
q_{req}=2.51 \times 10^{47} C (1+z)^3 \,{\rm s^{-1} \: Mpc^{-3}}.
\end{equation}
Here $C$ is the clumping factor of the IGM. Comparing the above 
expression for $q_{\rm req}$ to the rate of photon production by both stars and SNe, taking
into account our result from Table 1 that $0.12 f_{\rm esc,*} Q_{\rm *}
 \simeq f_{\rm esc,SN} Q_{\rm SN}$, we integrate over the galaxy luminosity function to find the following for the ratio of clumping factor to
escape fraction required for the universe to remain reionized:

\begin{equation}
\frac{C}{f_{\rm esc, *}}=1.12 \times 10^{-34} \frac{D_{\rm ion}}{(1+z)^3} \int
L_{1600}  \phi(L_{1600})  d L_{1600} \mbox{\ .}
\end{equation}
For the luminosity function $\phi$ we use the $z=7$ best-fit values
reported in Wilkins et al. (2011).  Given the uncertainty in the faint-end
slope of the luminosity function (e.g. Ryan et al. 2007; Khochfar et
al. 2007), we predict the ratio of $C/f_*$ for three different choices
of faint-end slopes  ($\alpha=-1.5, -1.7, -1.9; \phi^* = 0.00126,
0.00106,0.00072; M_{1600}^{*}=-19.8,-19.9,-20.1$).  Typical values
for the clumping factor and escape fraction are $C \sim 5$ (e.g. Pawlik
et al. 2009)  and $f_{\rm esc,*} \sim 0.2$ (e.g. Ciardi \& Ferrara 2005), respectively, which yields $C/f_{\rm esc,*}
\sim 25$.  As shown in Figure 4, such high values are only reached if
the luminosity function of galaxies is integrated down to a limiting
magnitude of $M_{1600}\sim -8$ and if the faint-end slope of the
luminosity function is as steep as $\alpha =-1.9$.  In the case of
only stellar photons the limiting magnitude is fainter by several
magnitudes, showing that the contribution to reionization from stars
in low luminous galaxies is comparable to the one from SNe in massive
galaxies.

Finally, we note that recent work suggests that the star formation rates of high redshift galaxies are not constant, as 
assumed in the calculation of the ionization parameter $D_{\rm ion}$ used above, but instead are increasing with time (e.g. Finlator et al. 2011; Khochfar \& Silk 2011).
In this event, $D_{\rm ion}$ should be considerably higher (Wilkins et al. 2011), and thus the contribution from low luminosity galaxies may be even 
lower than we find here.

\section{Discussion and Conclusions}
We have used state of the art models of stellar populations and high velocity shocks, along with the results of detailed radiative transfer simulations, to evaluate the contribution of SNe 
to the reionization of the universe.  We have found that, although the number of ionizing photons emitted from SN shocks is well below 
that emitted from stars, due to the harder spectrum of the ionizing radiation from SN shocks the fraction of such photons which escape galaxies is in general much larger than the escape fraction 
of stellar ionizing photons.  Related to this, owing principally to the softer spectrum of stars of higher metallicity, 
we find that the relative contribution of SNe to reionization is metallicity dependent; in particular, 
for stellar metallicities between 0.05 $Z_{\odot}$ $\la$ $Z$ $\la$ $Z_{\odot}$, SNe are responsible for between $\simeq$ 5 and $\simeq$ 
12 percent of the number of ionizations caused by stellar sources.   
Also, while these numbers are for line of sight-averaged escape fractions, we have furthermore found that, 
along a given line of sight out of a galaxy from a star-forming region, for hydrogen column densities $N_{\rm H}$ $\ga$ 10$^{18}$ cm$^{-2}$ the contribution 
to reionization from SNe is larger than that from stars.  Finally, we have shown that the contribution from SNe allows reionization to be completed by galaxies a few magnitudes brighter
than in the case in which only stellar sources contribute.

We emphasize that we have found a conservative estimate of the contribution from SNe, for the following reasons.  Firstly, we have included only the ionizing radiation emitted from SN shocks in the
Sedov-Taylor phase; while we find the radiated energy to already be $\simeq$ 40 percent of the total SN energy at this stage, there may be an additional contribution of ionizing photons
from SNe in the later, radiative phase if the upstream medium is already ionized by other sources (Shull \& Silk 1979; see also Section 2.1.1 here).  
Secondly, in our modeling we have not accounted 
for the fact that the majority of SN shock-generated photons are emitted after the stellar ionizing radiation from a given population of stars; as the radiation from the stars acts to ionize 
and rarify the gas along lines of sight out of a galaxy, when SNe explode the column density of neutral hydrogen $N_{\rm H}$ will likely be lower than the average column density seen by stellar photons.
We also note that, as we find that the number of ionizations caused by SN scales linearly with SN energy, if this is higher or lower than our fiducial value of $E_{\rm SN}$ = 10$^{51}$ erg during the 
epoch of reionization, then SNe may have contributed more or less, respectively, to reionization than we have found here.  Related to this, if the stellar IMF is more top-heavy during reionization than it 
is in the Galaxy today, as some recent studies would suggest (e.g. van Dokkum 2008; Dav{\' e} 2008; Gunawardhana et al. 2011), 
then this may also lead to an enhancement in the contribution from SNe, as in this case a higher fraction of stars may explode as SNe and these SNe may also be more energetic.
 
Beyond contributing to the ionization of hydrogen in the IGM, the enhanced photoheating rate due to the contribution of ionizing photons from SNe likely has additional effects.  
For instance, a higher heating rate of the gas contributes to the suppression of low-mass galaxy formation in reionized regions of the universe (e.g. Dijkstra et al. 2004; Hambrick et al. 2010).  
It is also likely to lead to a reduction in the clumping factor $C$ of the IGM, which has bearing on the rate at which reionization occurred; 
however, we note that for heating rates comparable to what would be expected for SN-generated ionizations 
Pawlik et al. (2009) find only a small decrease in the clumping factor.  

While we have found SNe to be of secondary importance compared to stars as contributors to reionization, they undeniably provide an enhancement in the 
rate at which reionization occurs above what is found in models accounting for only stellar sources (e.g. Sokasian et al. 2003; Iliev et al. 2006; 
Choudhury \& Ferrara 2007; Trac \& Cen 2007; Raicevi{\' c} et al. 2011).  In this, accounting for the effect of SNe can aid in explaining how the universe was reionized by $z$ $\sim$ 6.  
Using the results we have presented in Table 1, it is straightforward to include the contribution of SNe in reionization calculations that already include the contribution from stars.

\section*{Acknowledgements}
The authors are thankful to Jan-Pieter Paardekooper for many constructive comments on an early draft of this paper, as well as to Stephen Wilkins,
Umberto Maio, Fabrice Durier, Volker Gaibler, Bhaskar Agarwal, Eyal Neistein, Garrelt Mellema, Dan Whalen and Philipp Podsiadlowski for helpful discussions.  We would also 
like to acknowledge constructive feedback from an anonymous reviewer.


\begin{thebibliography}{1}

\bibitem[111(2000)]{a111}Allen, M.~G., Groves, B.~A., Dopita, M.~A., Sutherland, R.~S., Kewley, L.~J. 2008, ApJS, 178, 20
\bibitem[111(2000)]{a111}Alvarez, M., Abel, T. 2007, MNRAS, 380, 30
\bibitem[111(2000)]{a111}Barkana, R., Loeb, A. 2001, PhR, 349, 125
\bibitem[62(2000)]{a63}Blondin, J.~M., Wright, E.~B., Borkowski, K.~J., Reynolds, S.~P. 1998, ApJ, 500, 342
\bibitem[111(2000)]{a111}Bouwens, R.~J., et al. 2011, ApJ, submitted (arXiv:1105.2038)
\bibitem[111(2000)]{a111}Bowman, J.~D., Rogers, A.~E.~E. 2010, Nat, 468, 796
\bibitem[111(2000)]{a111}Bromm, V., Kudritzki, R.~P., Loeb, A. 2001, ApJ, 552, 464
\bibitem[111(2000)]{a111}Bromm, V., Yoshida, N. 2011, ARA\&A, in press (arXiv:1102.4638)
\bibitem[111(2000)]{a111}Bunker, A.~J., et al. 2010, MNRAS, 409, 855
\bibitem[1(2000)]{a1}Chevalier, R.~A. 1974, 188, 501
\bibitem[1(2000)]{a1}Chevalier, R.~A. 1977, ARA\&A, 15,175
\bibitem[1(2000)]{a1}Choudhury, T.~R., Ferrara, A. 2007, MNRAS, 380, L6
\bibitem[1(2000)]{a1}Ciardi, B., Ferrara, A. 2005, SSRv, 116, 625
\bibitem[1(2000)]{a1}Ciardi, B., Ferrara, A., White, S.~D.~M. 2003, MNRAS, 344, 7 
\bibitem[1(2000)]{a1}Cioffi, D.~F., McKee, C.~F., Bertschinger, E. 1988, ApJ, 334, 252
\bibitem[1(2000)]{a1}Clarke, C., Oey, M.~S. 2002, MNRAS, 337, 1299
\bibitem[1(2000)]{a1}Cox, D.~P. 1972, ApJ, 178, 159
\bibitem[1(2000)]{a1}Dav{\' e}, R. 2008, MNRAS, 385, 147
\bibitem[62(2000)]{a63}Dijkstra, M., Haiman, Z., Rees, M.~J., Weinberg, D.~H. 2004, ApJ, 601, 666
\bibitem[62(2000)]{a63}Dopita, M.~A., Krauss, L.~M., Sutherland, R.~S., Kobayashi, C., Lineweaver, C.~H. 2011, Ap\&SS, in press (arXiv:1106.5546)
\bibitem[62(2000)]{a63}Draine, B.~T., Woods, D.~T. 1991, ApJ, 383, 621 
\bibitem[1(2000)]{a1}Fan, X., Carilli, C.~L., Keating, B. 2006, ARA\&A, 44, 415
\bibitem[23(2000)]{a23}Finlator, K., Oppenheimer, B.~D., Dav{\' e}, R. 2011, MNRAS, 410, 1703
\bibitem[111(2000)]{a111}Gnedin, N.~Y., Ostriker, J.~P. 1997, ApJ, 486, 581
\bibitem[1(2000)]{a1}Gnedin, N.~Y., Kravtsov, A.~V., Chen, H.-W. 2008, ApJ, 672, 765
\bibitem[2(2000)]{a2}Greif, T.~H., Johnson, J.~L., Bromm, V., Klessen, R.~S. 2007, ApJ, 670, 1
\bibitem[23(2000)]{a23}Gunawardhana, M.~L.~P., et al. 2011, MNRAS, accepted (arXiv:1104.2379)
\bibitem[23(2000)]{a23}Haiman Z. 2009, Astrophysics in the Next Decade: JWST and Concurrent Facilities, Astrophysics\& Space Science Library, Eds. H. Thronson, A. Tielens, M. Stiavelli (arXiv:0809.3926)
\bibitem[23(2000)]{a23}Hambrick, D.~C., Ostriker, J.~P., Johansson, P.~H., Naab, T. 2010, MNRAS, submitted (arXiv:1009.6005)
\bibitem[23(2000)]{a23}Iliev, I.~T., Mellema, G., Pen, U.-L., Merz, H., Shapiro, P.~R., Alvarez, M.~A. 2006, MNRAS, 369, 1625
\bibitem[23(2000)]{a23}Jasche, J., Ciardi, B., En{\ss}lin, T.~A. 2007, MNRAS, 380, 417
\bibitem[23(2000)]{a23}Johnson, J.~L., Greif, T.~H., Bromm, V., Klessen, R.~S. Ippolito, J. 2009, MNRAS,399, 37
\bibitem[3(2000)]{a3}Khochfar, S., Silk, J. 2011, MNRAS, 410, L42
\bibitem[3(2000)]{a3}Khochfar, S., Silk, J., Windhorst, R.~A., Ryan, R.~E. 2007, ApJ, 668, L115
\bibitem[3(2000)]{a3}Kitayama, T., Yoshida, N. 2005, ApJ, 630, 675
\bibitem[3(2000)]{a3}Komatsu, E., et al. 2009, ApJS, 180, 330
\bibitem[3(2000)]{a3}Leitherer, C., et al. 1999, ApJS, 123, 3
\bibitem[4(2000)]{a4}Madau, P., Haardt, F., Rees, M.~J. 1999, ApJ, 514, 648
\bibitem[4(2000)]{a4}Mansfield, V.~N., Salpeter, E.~E. 1974, ApJ, 190, 305
\bibitem[4(2000)]{a4}McKee, C.~F., Tan, J.~C. 2008, ApJ, 681, 771
\bibitem[4(2000)]{a4}McQuinn, M., Lidz, A., Zahn, O., Dutta, S., Hernquist, L., Zaldarriaga, M. 2007, MNRAS, 377, 1043
\bibitem[4(2000)]{a4}Miniati, F., Ferrara, A., White, S.~D.~M., Bianchi, S. 2004, MNRAS, 348, 964
\bibitem[4(2000)]{a4}Mirabel, I.~F., Dijkstra, M., Laurent, P., Loeb, A., Pritchard, J.~R. 2011, A\&A, 528, 149
\bibitem[4(2000)]{a4}Miralda-Escud{\' e}, J., Haehnelt, M., Rees, M.~J. 2000, ApJ, 530, 1
\bibitem[23(2000)]{a23}Nagashima, M., Lacey, C.~G., Baugh, C.~M., Frenk, C.~S., Cole, S. 2005, MNRAS, 358, 1247
\bibitem[5(2000)]{a5}Oh, S.~P. 2001, ApJ, 553, 499
\bibitem[23(2000)]{a23}Osterbrock, D.~E., Ferland, G.~J. 2006, Astrophysics of gaseous nebulae and active galactic nuclei, University Science
\bibitem[5(2010)]{a67}Ouchi, M., et al. 2009, 706, 1136
\bibitem[23(2000)]{a23}Paardekooper, J.-P., Pelupessy, F.~I., Altay, G., Kruip, C.~J.~H. 2011, A\&A, submitted
\bibitem[23(2000)]{a23}Pawlik, A.~H., Schaye, J., van Scherpenzeel, E. 2009, MNRAS, 394, 1812
\bibitem[23(2000)]{a23}Power, C., Wynn, G.~A., Combet, C., Wilkinson, M.~I. 2009, MNRAS, 395, 1146
\bibitem[1(2000)]{a1}Raicevi{\' c}, M., Theuns, T., Lacey, C., 2011, MNRAS, 410, 775
\bibitem[23(2000)]{a23}Razoumov, A.~O., Sommer-Larsen, J. 2010, ApJ, 710, 1239
\bibitem[23(2000)]{a24}Ricotti, M., Ostriker, J.~P. 2004, MNRAS, 352, 547
\bibitem[1(2000)]{a1}Robertson, B.~E., Ellis, R.~S., Dunlop, J.~S., McLure, R.~J., Stark, D.~P. 2010, Nat, 468, 49
\bibitem[23(2000)]{a23}Ruderman, M.~A. 1974, Sci, 184, 1079
\bibitem[1(2000)]{a1}Ryan, R.~E., et al. 2007, ApJ, 668, 839
\bibitem[62(2000)]{a62}Schaerer, D. 2002, A\&A, 382, 28
\bibitem[62(2000)]{a62}Schleicher, D.~R.~G., Banerjee, R., Klessen, R.~S. 2008, PhRvD, 78, 3005
\bibitem[62(2000)]{a63}Shin, M.-S., Trac, H., Cen, R. 2008, ApJ, 681, 756
\bibitem[62(2000)]{a64}Shu, F.~H. 1992, The Physics of Astrophysics, University Science
\bibitem[62(2000)]{a64}Shull, J.~M., McKee, C.~F. 1979, ApJ, 227, 131
\bibitem[6(2000)]{a6}Shull, J.~M., Silk, J. 1979, ApJ, 234, 427
\bibitem[7(2000)]{a7}Shull, J.~M., van Steenberg, M.~E. 1985, ApJ, 298, 268
\bibitem[1(2000)]{a1}Sokasian, A., Abel, T., Hernquist, L., Springel, V. 2003, MNRAS, 344, 607
\bibitem[23(2000)]{a23}Spitzer, L., Tomasko, M.~G. 1968, ApJ, 152, 971
\bibitem[7(2000)]{a7}Stacy, A., Bromm, V. 2007, MNRAS, 382, 229
\bibitem[62(2000)]{a63}Sutherland, R.~S., Dopita, M.~A. 1993, ApJS, 88, 253
\bibitem[7(2000)]{a7}Tegmark, M., Silk, J., Evrard, A. 1993, ApJ, 417, 54
\bibitem[1(2000)]{a1}Trac, H., Cen, R. ApJ, 671, 1
\bibitem[7(2000)]{a7}Truelove, J.~K., McKee, C.~F. 1999, ApJS, 120, 299
\bibitem[23(2000)]{a23}van Dokkum, P.~G. 2008, ApJ, 674, 29
\bibitem[8(2000)]{a8}Volonteri, M., Gnedin, N.~Y. 2009, ApJ, 703, 2113
\bibitem[9(2000)]{a9}Whalen, D., van Veelen, B., O’Shea, B.~W., Norman, M.~L. 2008, ApJ, 682, 49
\bibitem[23(2000)]{a23}Wilkins, S.~M., Bunker, A.~J., Lorenzoni, S., Caruana, J. 2011, MNRAS, 411, 23
\bibitem[10(2000)]{a10}Wise, J.~H., Cen, R. 2009, ApJ, 693, 984
\bibitem[10(2000)]{a10}Wyithe, J.~S.~B., Hopkins, A.~M., Kistler, M.~D., Y{\" u}ksel, H., Beacom, J.~F. 2010, MNRAS, 401, 2561 
\bibitem[23(2000)]{a23}Yajima, H., Choi, J.-H., Nagamine, K. 2011, MNRAS, 412, 411
\bibitem[23(2000)]{a23}Yajima, H., Umemura, M., Mori, M., Nakamoto, T. 2009, MNRAS, 398, 715
\bibitem[23(2000)]{a23}Yamada, M., Nishi, R. 1998, ApJ, 505, 148
\bibitem[23(2000)]{a23}Yoshida, N., Bromm, V., Hernquist, L. 2004, ApJ, 605, 579

\end{thebibliography}
\bibliographystyle{apj}

\appendix

\section{An approximate expression for the escape fraction along a single line of sight}
Here we justify the expression that we have used for the fraction of ionizing photons which escape along a single line of sight, 
equation (13).  The exact expression for the escape fraction of ionizing photons along a given line of sight with column density of
hydrogen $N_{\rm H}$ is given by

\begin{equation}
f_{\rm esc, *} = \frac{\int_{{\rm 13.6 eV}}^{\infty} \frac{F(E_{\gamma})}{E_{\gamma}}{\rm e}^{-N_{\rm H} \sigma_{\rm *}(E_{\rm \gamma})} dE_{\gamma}}{\int_{{\rm 13.6 eV}}^{\infty} \frac{F(E_{\gamma})}{E_{\gamma}} dE_{\gamma}} \mbox{\ ,}
\end{equation}
where for concreteness we have chosen to treat the case of stellar  photons; a similar treatment to that given 
here can easily be given for the case of SNe  photons.  In order to show that equation (13) is approximately equal to this expression,
we expand the exponential in the numerator, which yields

\begin{equation}
f_{\rm esc, *} \simeq \frac{\int_{{\rm 13.6 eV}}^{\infty} \frac{F(E_{\gamma})}{E_{\gamma}}\left[1 - N_{\rm H}\sigma_{\rm *}(E_{\rm \gamma})\right] dE_{\gamma}}{\int_{{\rm 13.6 eV}}^{\infty} \frac{F(E_{\gamma})}{E_{\gamma}} dE_{\gamma}} \mbox{\ .}
\end{equation}
Slightly rewriting this, we have 

\begin{equation}
f_{\rm esc, *} \simeq 1 - N_{\rm H} \frac{\int_{{\rm 13.6 eV}}^{\infty} \frac{F(E_{\gamma})}{E_{\gamma}} \sigma_{\rm *}(E_{\rm \gamma}) dE_{\gamma}}{\int_{{\rm 13.6 eV}}^{\infty} \frac{F(E_{\gamma})}{E_{\gamma}} dE_{\gamma}} \mbox{\ .}
\end{equation}
Noting that the cross section here is that due to absorption of ionizing photons by both hydrgen and helium, as discussed in Section 3.1, 
we have $\sigma_{\rm *}(E_{\rm \gamma}) = \sigma_{\rm HI}(E_{\gamma}) + \sigma_{\rm HeI}(E_{\gamma})$.  Then, with equation (12), we obtain

\begin{equation}
f_{\rm esc, *} \simeq 1 - N_{\rm H}\sigma_{\rm *} \simeq {\rm e}^{-N_{\rm H}\sigma_{\rm *}} \mbox{\ ,}
\end{equation}
which is equation (13) for the case of stellar photons.  This demonstrates that  equation (13)
is valid as an approximate expression for the escape fraction.

\end{document}